\documentclass[preprint,aps,showpacs,nofootinbib,preprintnumbers,amsmath,amssymb]{revtex4-1}
\usepackage{epsfig}
\usepackage{subfigure}
\usepackage{multirow}
\usepackage{dcolumn}
\usepackage{bm}
\usepackage[usenames ,dvipsnames]{xcolor}
\usepackage{slashed}
\usepackage{float}
\usepackage{subfigure}
\usepackage{graphicx,color}
\usepackage{amsmath}
\usepackage{slashed}
\usepackage{nonfloat}
\begin{document}
\title{Revisiting semileptonic  $B^-\to p\bar{p} \ell^- \bar{\nu}_{\ell}$  decays}

\author{C.Q. Geng$^{1,2,3}$, Chia-Wei Liu$^{2}$ and Tien-Hsueh Tsai$^{3}$}
\affiliation{
$^{1}$Chongqing University of Posts \& Telecommunications, Chongqing 400065, China\\
$^{2}$School of Fundamental Physics and Mathematical Sciences, Hangzhou Institute for Advanced Study, UCAS, Hangzhou 310024, China 
\\
$^{3}$Department of Physics, National Tsing Hua University, Hsinchu 300, Taiwan
}\date{\today}

\begin{abstract}
We systematically revisit the baryonic four-body semileptonic decays of $B^- \to {\bf B}\bar{\bf B}'\ell^- \bar{\nu}_{\ell}$ by the perturbative QCD counting rules with ${\bf B}$ representing octet baryons and $\ell=e,\mu$.  We study  the  transition form factors of $B^- \to {\bf B} \bar{\bf B}'$ in the limit of 
$(p_{\bf B}+p_{\bar{\bf B}'})^2 \to \infty $  with  the  three-body $\bar{B}\to {\bf B}\bar{\bf B}' M$ and $B^- \to p\bar{p} \mu^- \bar{\nu}_{\mu}$ data along with $SU(3)_f$ flavor symmetry. We calculate the decay branching ratios and angular asymmetries  as well as  the differential decay branching fractions of $B^- \to p \bar{p} \ell^- \bar{\nu}_{\ell}$.
In particular, we find that our new result of ${\cal B}( B^- \to p \bar{p} \ell^- \bar{\nu}_{\ell})=(5.21\pm0.34)\times 10^{-6}$, which is about one order of magnitude lower than the previous theoretical prediction of $(10.4\pm2.9)\times 10^{-5}$, agrees well with both experimental  measurements of  
$(5.8^{+2.6}_{-2.3})\times 10^{-6}$ and $(5.3\pm0.4)\times 10^{-6}$ by the Belle and LHCb Collaborations, respectively. We also evaluate the branching ratios and angular asymmetries in other channels of  $B^- \to {\bf B}\bar{\bf B}\ell^- \bar{\nu}_{\ell}$,
which can be tested by the ongoing experiments at  LHCb and  BelleII. 

\end{abstract}
\maketitle

\section{introduction}
In 2011, the baryonic four-body semileptonic decay of  $B^-\to p\bar{p} \ell^- \bar{\nu}_{\ell}$ ($\ell=e$ or $\mu$) was studied
with its decay branching ratio predicted to be $(10.4\pm 2.6\pm 1.2)\times 10^{-5}$ in Ref.~\cite{Geng:2011tr}.
 Both $e$ and $\mu$ modes were indeed measured by the Belle Collaboration~\cite{Tien:2013nga} in 2014 with 
 ${\cal B}(B^- \to p \bar{p} e^- \bar{\nu}_{e})=(8.2^{+3.7}_{-3.2}\pm0.6)\times 10^{-6}$ and 
  ${\cal B}(B^- \to p \bar{p} \mu^- \bar{\nu}_{\mu})=(3.1^{+3.1}_{-2.4}\pm0.7)\times 10^{-6}$
   along with the combined value of ${\cal B}( B^- \to p \bar{p} \ell^- \bar{\nu}_{\ell})=(5.8^{+2.4}_{-2.1}\pm0.9)\times 10^{-6}$,
  which is  about one oder of magnitude lower than the theoretical prediction.
Recently, the LHCb Collaboration has published the observation of   $B^-\to p \bar{p} \mu^- \bar{\nu}_{\mu}$
with its decay branching ratio  determined to be   $(5.27^{+0.23}_{-0.24}\pm 0.21\pm0.15)\times 10^{-6} $~\cite{Aaij:2019bdu}, 
where the first and second uncertainties correspond to  statistical and systematic, and the third one is from  the branching fraction of the normalization channel,
respectively. 
Both statistical and systematic uncertainties of the LHCb data
 have significant improvements compared with the previous ones by Belle~\cite{Tien:2013nga}.
 
 These decay modes are useful to determine the value of $|V_{ub}|$ as the works in the other baryonic modes\cite{Hsiao:2018zqd} as well as the underlying  new CP/T violating effects. The main difficulty in both extracting $|V_{ub}|$ from $B^- \to p\bar{ p}\ell \bar{\nu}_{\ell}$ and constraining the new CP/T violating effects is how to obtain their hadronic transition amplitude of $B^- \to p \bar{p}$
as it is hard to be calculated via the usual QCD methods, such as the 
factorizations and  sum rules, which have been widely used in the mesonic decays 
of $B^- \to \pi^+ \pi^- \ell^- \bar{\nu}_{\ell}$ ($B_{\ell 4}$)~\cite{Feldmann:2018kqr,Boer:2016iez,Cheng:2017smj}.
 Nevertheless, these modes should be considered in the fit for the extraction of $V_{ub}$. Qualitatively speaking,
 to reduce the theoretical values for the decay branching ratios of $B^- \to p\bar{ p}\ell \bar{\nu}_{\ell}$, a smaller value of $|V_{ub}|$ is needed 
 besides the form factors. It is similar to the extractions from the exclusive $B$ and $\Lambda_b$ decays, but lower than that from the inclusive B decays.
  Clearly,  as a baryonic complementary version of $B_{\ell 4}$ decays, both  theoretical and experimental studies of $B^- \to p \bar{p} \ell^- \bar{\nu}_{\ell}$ may shed light on the baryonic transition amplitude of $B^- \to p \bar{p}$, uncover the nature of the QCD dynamics, 
  and improve the measurement of $|V_{ub}|$.

\begin{figure}[h]
 	\begin{minipage}[h]{0.4\linewidth}
 		\centering
 		\includegraphics[width=3in]{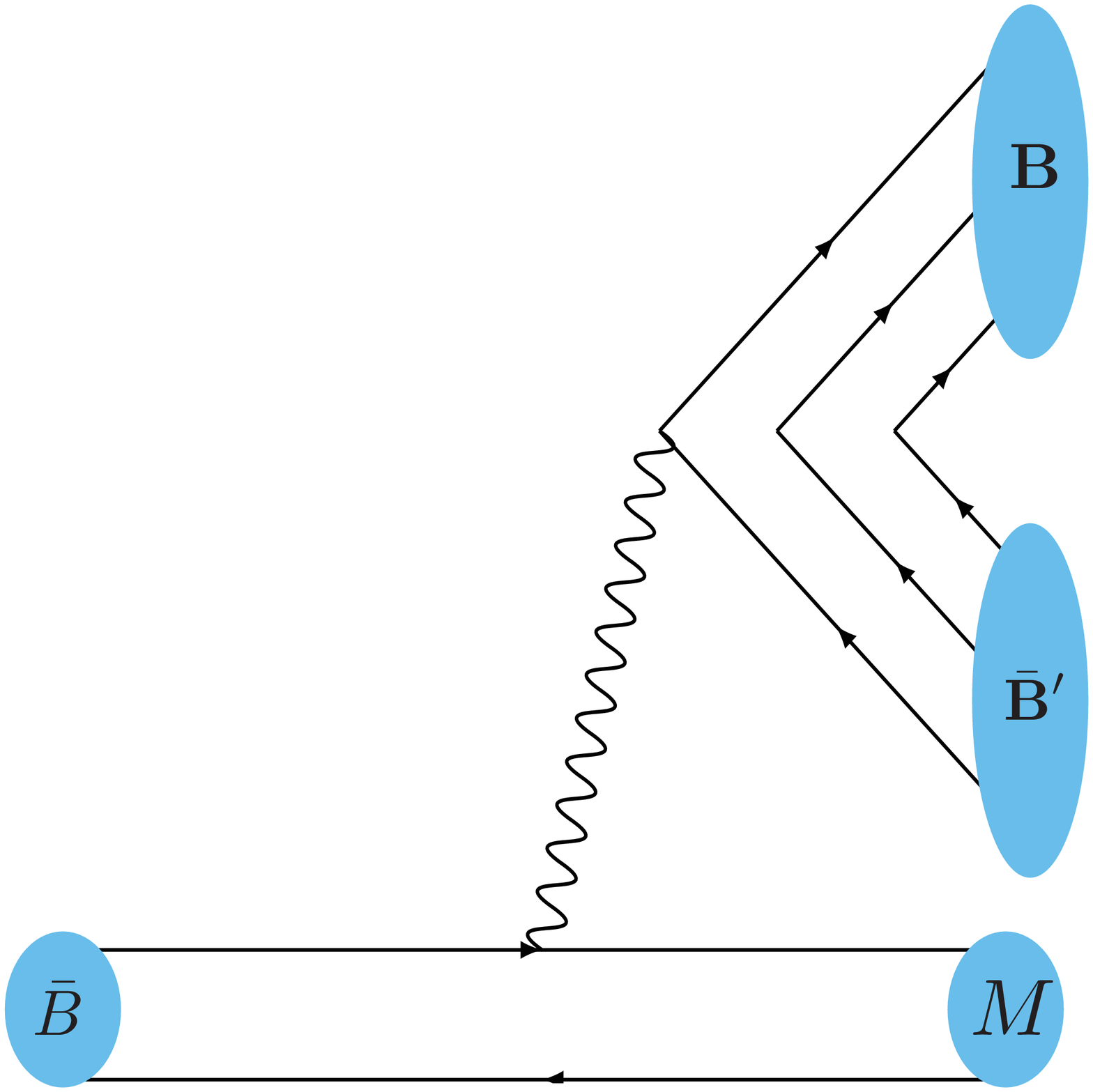}
 		\\
 		(a)
 	\end{minipage}
 	\begin{minipage}[h]{0.4\linewidth}
 		\centering
 		\includegraphics[width=2.6in]{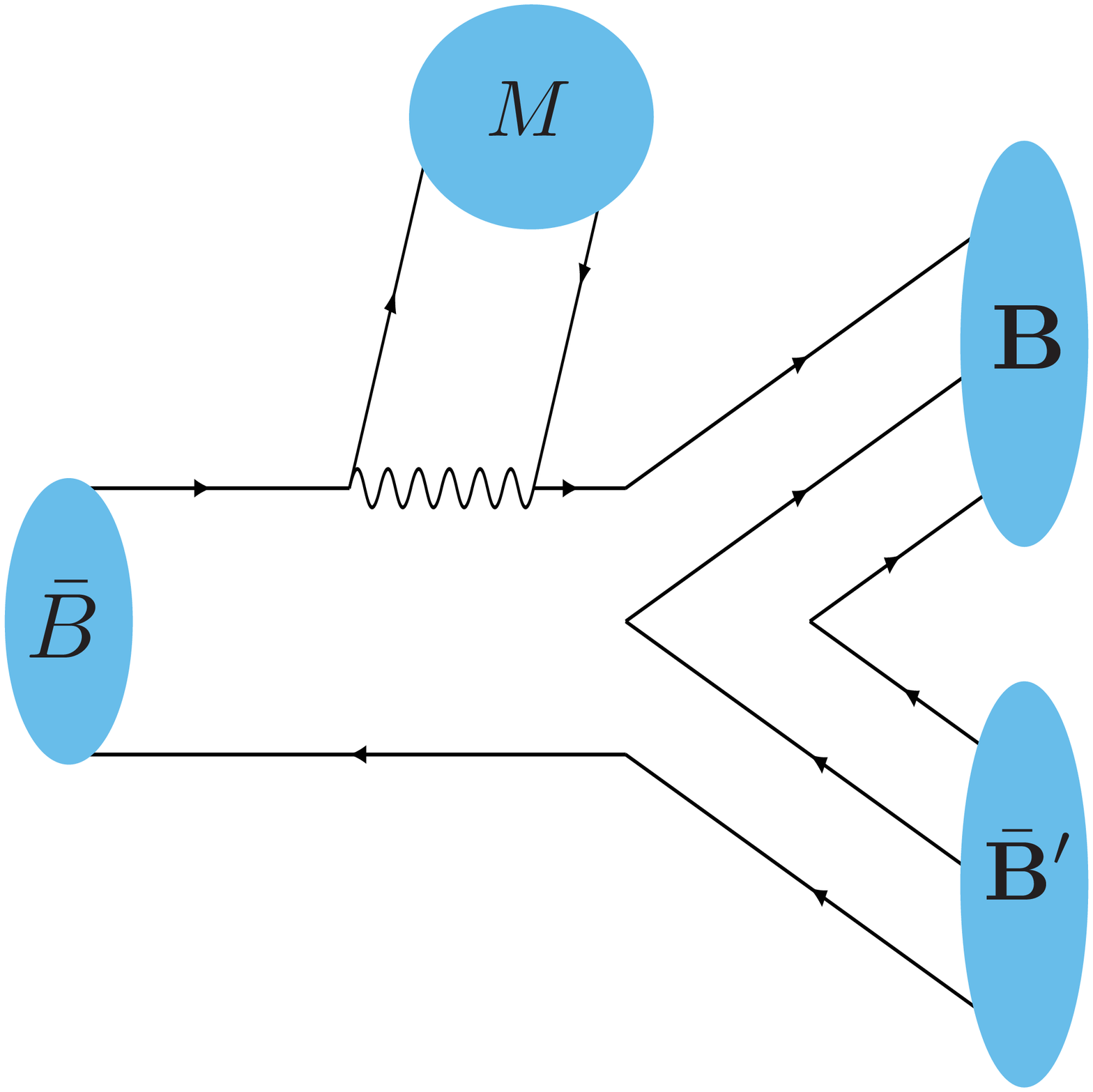}
 		\\
 		(b)
 	\end{minipage}
 	\caption{Quark flow diagrams for  three-body baryonic B decays $\bar{B} \to {\bf B}\bar{\bf B}' M$ with (a)  current and (b) transition types.}
	\label{F1}
 \end{figure}

Because of the rareness of four-body $B^-  \to {\bf B}\bar{\bf B}' \ell^- \bar{\nu}_{\ell}$  decays with ${\bf B}^{(\prime)}$ representing octet baryons,
 people have concentrated  on the three-body $\bar{B}\to  {\bf B}\bar{\bf B}'  M$ decays
  to extract the baryonic transition from factors in the ${\bf B}\bar{\bf B}'$ transitions,
  where ${\bf B}(\bar{\bf B}')$ and $M$ are octet (anti-)baryons and    pseudoscalar or vector mesons, respectively.
There have been several theoretical analyses on the baryonic three-body $B\to {\bf B} \bar{\bf B}' M$ decays based on the factorization assumptions~\cite{Cheng:2001tr,Cheng:2001ub,Cheng:2002fp,Chua:2002yd,Chua:2002wn,Geng:2005fh,Geng:2005fh,Chen:2008sw}.
These  baryonic $B$ decays can be basically classified into  current production $\cal C$, transition $\cal T$ and  hybrid ${\cal C}+{\cal T}$
types~\cite{Chen:2008sw}, with the quark flow diagrams shown in Fig.~\ref{F1}.
 Among them, the transition one is the only channel directly related to the $\bar{B} \to {\bf B}\bar{\bf B}' $ baryonic transition amplitudes. In Ref.~\cite{Geng:2011tr}, perturbative QCD counting rules combined with the available data of $\bar{B} \to {\bf B}\bar{\bf B}' M$ decays at the time
 were used to fit the form factors and predict
  the $B^- \to p \bar{p} \ell^- \bar{\nu}_{\ell}$ decays.  
  Although the prediction of $B^- \to p \bar{p} \ell^- \bar{\nu}_{\ell}$ motivated its active search, it is clearly disproved by the experiments of 
  both Belle~\cite{Tien:2013nga} and LHCb~\cite{Aaij:2019bdu}.
  The main reason for such a large prediction is that there were short of relevant data  as well as lack of the understanding of the
  underlined QCD dynamics for the baryonic transition of $B^- \to p \bar{p}$. 
In this work, we would like to
  reanalyze the semi-leptonic decays of $B^- \to p \bar{p} \ell^- \bar{\nu}_{\ell}$ 
  with the same strategy as that in Ref.~\cite{Geng:2011tr} with the updated data. In addition, we 
  shall use the flavor symmetry to extend our results to other $B^- \to {\bf B} \bar{\bf B} \ell^- \bar{\nu}_{\ell}$ decays. 
This work is the first step to know the property of ${\bar B} \to {\bf B}{\bf B}'$ transition matrix elements. After getting a better understanding of these elements, we can use them for not only improving the measurement of $|V_{ub}|$ but probing or constraining the new physics effects, such as the T violating  triple momentum correlations due to the rich kinematic structure in the four-body decays of ${\bar B} \to {\bf B}{\bf B}' \ell \bar{\nu}$.  

 This paper is organized as follows.  In Sec.~II, we present our formalism, which contains the effective Hamiltonians and generalized transition form factors.
 In Sec. III, we show our numerical results of the form factors fitted by three-body $\bar{B}\to {\bf B}\bar{\bf B}' M$ processes and the latest $B^-\to p \bar{p}\mu^-\bar{\nu}_{\mu}$ result,
 and  present our predictions of the branching ratios and angular asymmetries in $B^- \to {\bf B} \bar{\bf B} \ell^- \bar{\nu}_{\ell}$.
 We also compare our results of  $p\bar{p}$ invariant mass spectrum in the $B^- \to p \bar{p} \mu^- \bar{\nu}_{\mu}$ decay with the one measured by the LHCb. We give our  conclusions in Sec.~IV.
 
\section{Formalism}
\begin{figure}
	\includegraphics[width=3in]{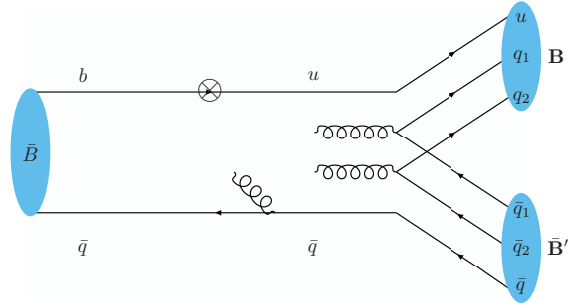}
	\caption{Diagram for the $\bar{B} \to \bar{\bf B} {\bf B}'$ transition, where the curl lines stand for hard-gluons and  the symbol of $\otimes$ denotes the weak vertex, while each hard gluon contributes an $1/t$ in the form factors.}
	\label{F2}
\end{figure}

The effective Hamiltonian for $\bar{B}\to {\bf B}\bar{\bf B}'  \ell^- \bar{\nu}_{\ell}$ at the quark level is given by
\begin{eqnarray} \label{effH}
{\cal H}_{eff}=\frac{G_F}{\sqrt{2}}V_{ub}\bar{u}\gamma^{\mu}(1-\gamma_5)b{\bar \ell }\gamma_\mu(1-\gamma_5)\nu_{\ell}\,,
\end{eqnarray}
where $G_F$ is the Fermi constant and $V_{ub}$ represents the element of the CKM matrix. 
The transition amplitude of $\bar{B}\to {\bf B}\bar{\bf B}'  \ell^- \bar{\nu}_{\ell}$ can be easily factorized into  hadronic  and  leptonic parts,  written as
\begin{eqnarray}
	{\cal A}(\bar{B}\to {\bf B}\bar{\bf B}'  \ell^- \bar{\nu}_{\ell})=\frac{G_F}{\sqrt{2}}V_{ub}\langle {\bf B}\bar{\bf B}'|\bar{u}\gamma^\mu(1-\gamma_5)|\bar{B}\rangle \bar{\ell}\gamma_\mu(1-\gamma_5)\nu_{\ell}\,,
\end{eqnarray}
where $\bar{\ell}$ and $\nu_{\ell}$ are the usual Dirac spinors and $\langle {\bf B}\bar{\bf B}'|\bar{u}\gamma^\mu(1-\gamma_5)|\bar{B}\rangle$ is the unknown hadronic transition amplitude. The most general Lorentz invariant forms of the hadronic  transitions for the vector and axial-vector currents
 can be parametrized by~\cite{Geng:2011tr,Chen:2008sw}
\begin{eqnarray}
	\langle {\bf B}\bar{\bf B}'|\bar{u}\gamma^\mu b|\bar{B}\rangle=i\bar{u}(p_{\bf B})\left[ g_1\gamma^\mu+ig_2\sigma^{\mu \nu}p_{\nu}+g_3 p^{\mu} +g_4(p_{\bf B}+p_{\bar{\bf B}'})^{\mu}+g_5(p_{\bf B}-p_{\bar{\bf B}'})^{\mu}\right]\gamma_5 v(p_{\bar{\bf B}'}),\nonumber \\
		\langle {\bf B}\bar{\bf B}'|\bar{u}\gamma^\mu \gamma_5 b|\bar{B}\rangle=i\bar{u}(p_{\bf B})\left[ f_1\gamma^\mu+if_2\sigma^{\mu \nu}p_{\nu}+f_3 p^{\mu} +f_4(p_{\bf B}+p_{\bar{\bf B}'})^{\mu}+f_5(p_{\bf B}-p_{\bar{\bf B}'})^{\mu}\right] v(p_{\bar{\bf B}'}),~~\, 
\end{eqnarray} 
respectively,
where $f_i$ and $g_i$ ($i=1,2,\cdots,5$) are the form factors and  $p^\mu=(p_{B^-}-p_{\bf B}-p_{\bar{\bf B}'})^\mu$. 
Inspiring from the threshold effects~\cite{HS}, which have been observed in three-body $\bar{B}\to {\bf B} \bar{\bf B}' M$  decays~\cite{Aaij:2017vnw,Aubert:2006qx,Abe:2002tw}, and the pQCD counting rules~\cite{Brodsky:1980sx,Lepage:1980fj,Lepage:1979za}, the momentum dependences of $f_i$ and $g_i$ can be assumed to be
\begin{eqnarray}
\label{FF}
	f_i=\frac{C_{f_i}}{t^n}, \quad g_i=\frac{C_{g_i}}{t^n},
\end{eqnarray}  
with $n=3$, where $C_{f_i,g_i}$ are constants determined by the branching ratios of the input channels.
  Note that $n$ relates to the number of  hard-gluon propagators as shown in Fig.~\ref{F2}.  In the $\bar{B}\to {\bf B}\bar{\bf B}'$ transition, two hard gluons produce the valance quarks in the ${\bf B}\bar{\bf B}'$ pair separately as well as one more hard gluon is needed to speed up the spectator quark in $\bar{B}$~\cite{Chen:2008sw}. As a result, we can use   $C_{f_i}$ and $C_{g_i}$ to describe the hadronic form factors in both transition type three-body and semileptonic four-body decays.
With the help of  $SU(3)$ flavor and $SU(2)$ spin symmetries in $t \to \infty$ and heavy quark limit,  $C_{f_i}$ and $C_{g_i}$ are related by only  two chiral-conserving parameters $C_{RR}$ and $C_{LL}$  and one chiral-flipping parameter $C_{LR}$. Consequently, we have
\begin{eqnarray}
 C_{f_1}&=&m_{\bar{B}}\left(e_{LL}C_{LL}+e_{RR}C_{RR}\right)+\left(m_{\bf B}+m_{\bar{\bf B}'}\right)e_{LR}C_{LR}\nonumber\\
 C_{g_1}&=&m_{\bar{B}}\left(e_{LL}C_{LL}-e_{RR}C_{RR}\right)+\left(m_{\bf B}-m_{\bar{\bf B}'}\right)e_{LR}C_{LR}\nonumber\\
 C_{f_2}&=&-C_{g_2}=e_{LR}C_{LR},\quad C_{f_i}=C_{g_i}=-e_{LR}C_{LR}\quad \text{for }i=3,4,5\,,
 \label{relation}
\end{eqnarray}
where $e_{RR}$, $e_{LL}$ and $e_{LR}$ are the electroweak coefficients determined by the spin-flavor structure of $\bar{B}$ and ${\bf B}\bar{\bf B}'$,
and  $m_{{\bf B},\bar{\bf B}'}$ correspond to baryon and
anti-baryon masses, respectively. The detail derivations of Eq.~\ref{relation} are presented in Appendix.
We list the coefficients of relevant channels in Table~\ref{chiral}. 
\begin{table}[t]
	\caption{ Electroweak coefficients of $\bar{B}\to {\bf B}\bar{\bf B}'$  under $t\to \infty$ and heavy quark limits.}
	\begin{tabular}{c|c|c|c}
		\hline
		\hline
		Channel& $e_{RR}$&$e_{LL}$&$e_{LR}$\\
		\hline
		$B^-\to p \bar{p}$&$\frac{1}{3}$&$\frac{5}{3}$&$-\frac{4}{3}$\\
		$B^- \to n \bar{n}$&$\frac{2}{3}$&$\frac{1}{3}$&$\frac{1}{3}$\\
		$B^- \to \Sigma^+ \bar{\Sigma}^+$&$\frac{1}{3}$&$\frac{5}{3}$&$-\frac{4}{3}$\\
		$B^- \to \Sigma^0 \bar{\Sigma}^0$&$\frac{1}{6}$&$\frac{5}{6}$&$-\frac{2}{3}$\\
		$B^- \to \Xi^0 \bar{\Xi }^0$&$\frac{2}{3}$&$\frac{1}{3}$&$\frac{1}{3}$\\
		$B^- \to \Lambda^0 \bar{\Lambda}^0$&$\frac{1}{2}$&$\frac{1}{3}$&$0$\\
		$B^- \to \Lambda^0 \bar{\Sigma}^0$&$-\frac{1}{2\sqrt{3}}$&$\frac{1}{2\sqrt{3}}$&$-\frac{1}{\sqrt{3}}$\\
		$B^- \to \Sigma^0 \bar{\Lambda}^0$&$-\frac{1}{2\sqrt{3}}$&$\frac{1}{2\sqrt{3}}$&$-\frac{1}{\sqrt{3}}$\\
		$B^- \to \Lambda^0 \bar{p}$&$0$&$\sqrt{\frac{2}{3}}$&$-\sqrt{\frac{2}{3}}$\\
		$\bar{B}^0 \to \Lambda^0 \bar{\Lambda}^0$&$\frac{1}{2}$&$\frac{1}{2}$&$0$\\
		$\bar{B}^0\to p \bar{p}$&$\frac{2}{3}$&$\frac{1}{3}$&$\frac{1}{3}$\\
		\hline
	\end{tabular}
\label{chiral}
\end{table}

\begin{figure}
		\includegraphics[width=4in]{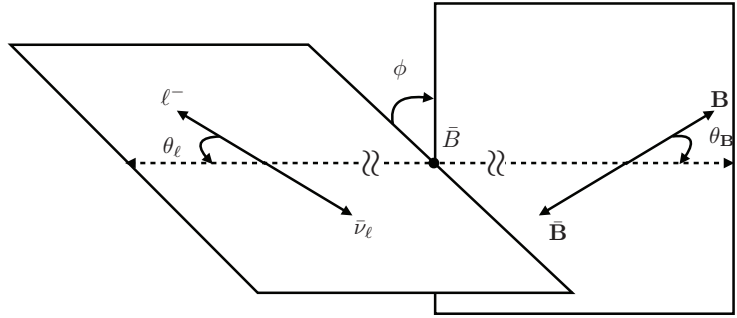}
	\caption{$\theta_{\bf B},\theta_{\ell}$ and $\phi$ in $\bar{B}\to {\bf B}\bar{\bf B}\ell^- \bar{\nu}_{\ell}$ decays.}
	\label{kin}
\end{figure}

Following the same formalism in the literature of $B_{\ell 4},D_{\ell 4}$ and $K_{\ell 4}$ analyses~\cite{Lee:1992ih,Pais:1968zza}, 
we examine the $\bar{B}\to {\bf B}\bar{\bf B}' \ell^- \bar{\nu}_{\ell}$  system in the $\bar{B}$  rest frame with five kinematic variables,  $s=(p_{\ell}+p_{\bar{\nu}_{\ell}})^2,~t,~\theta_{\bf B},~\theta_{\ell}$ and $\phi$, 
where $\sqrt{s}$ and $\sqrt{t}$ are the invariant masses of lepton and ${\bf B}\bar{\bf B}'$ pairs, respectively, and three kinematic angles are shown in Fig.~\ref{kin}.
The differential decay width is given by
\begin{eqnarray}
	d\Gamma=\frac{|\bar{\cal A}|^2}{4(4\pi)^6m_{\bar{B}}^3}X\beta_{\bf B}\beta_{\bf L}ds dt d\cos\theta_{\bf B} d\cos\theta_{\ell}d\phi\,,
\end{eqnarray}
where $|\bar{\cal A}|^2$ is the spin-averaged amplitude and $X,\beta_{\bf B},\beta_{\bf L}$ are given by
\begin{eqnarray}
	X&=&\frac{\sqrt{(m_{\bar{B}}^2-s-t)^2-4st}}{2}\nonumber\\
	\beta_{\bf B}&=&\frac{1}{t}\lambda^{\frac{1}{2}}(t,m_{\bf B}^2,m_{\bar{\bf B}'}^2)\nonumber\\
	\beta_{\bf L}&=&\frac{1}{s}\lambda^{\frac{1}{2}}(s,m_{\ell}^2,m_{\bar{\nu}}^2)\,
\end{eqnarray}
with $\lambda(x,y,z)=x^2+y^2+z^2-2xy-2xz-2yz$. 
We can also define the integrated $\theta_{\bf B}$ and $\theta_{\ell}$ asymmetries of  ${\bf B}\bar{\bf B}'$  and   lepton pairs as follows:
\begin{eqnarray}
	\langle\alpha_{\theta_{f}}\rangle&\equiv&\frac{\int_{0}^{1}\frac{d\Gamma}{d\cos\theta_{f}}d\cos\theta_{f}
	-\int_{-1}^{0}\frac{d\Gamma}{d\cos\theta_{f}}d\cos\theta_{f}}{\int_{0}^{1}\frac{d\Gamma}{d\cos\theta_{f}}d\cos\theta_{f}+\int_{-1}^{0}\frac{d\Gamma}{d\cos\theta_{f}}d\cos\theta_{f}}\,,
\end{eqnarray}
 with $f={\bf B}$ and $\ell$, respectively.

\section{Numerical Results}
In our numerical analysis, we use the Wolfenstein parameterization for the CKM matrix with the corresponding parameters, taken to be~\cite{pdg}
\begin{eqnarray}
	&\lambda=0.22650,\quad A=0.790, \quad \rho=0.141,\quad \eta=0.357,
\end{eqnarray}
leading to $|V_{ub}|=3.8\times 10^{-3}$.
To extract the  form factors, we use the factorization assumption and follow the formula in Ref.~\cite{Chen:2008sw} to calculate the branching ratios of $\bar{B}\to {\bf B}\bar{\bf B}' M$. The full analysis of three-body kinematics and detail derivations of $\bar{B}\to {\bf B}\bar{\bf B}' M$ factorization amplitudes can be found in Ref.~\cite{Chen:2008sw}.  Based on Refs.~\cite{Chen:2008sw,Cheng:1998kd,Cheng:2001sc}, the effective Wilson coefficients of $a_2^{D^{(*)}}$ and $a_2^{J^{(*)}}$ should include non-factorizable effects and can be parameterized by the effective color number $(N_c^{eff})$, $a_2^{M}=c_2^M+c_1^M/N_c^{eff}$, where $N_c^{eff}$ will be fitted .
We present the numerical inputs of hadron masses, lifetimes, meson decay constants, and  Wilson coefficients 
in Table~\ref{input}~\cite{Geng:2011tr,pdg,Buchalla:1995vs}.
\begin{table}[t]
	\caption{Input values of hadron masses, lifetimes, meson decay constants, and  Wilson coefficients, 
	where the masses (decay constants) and lifetimes  are in  units of GeV and fs, while the Wilson coefficients are dimensionless.}
	\begin{tabular}{cccccccccc}
		\hline
		\hline
		$m_{B^-}$&$m_{B^0}$&$m_{D}$&$m_{D^*}$&$m_{J/\Psi}$&$m_{p}$&$m_{n}$&$m_{\Lambda^0}$&$m_{\Xi^0}$&	$m_{\Sigma^0}$\\
		\hline
		$5.28$&$5.28$&$1.87$&$2.01$&$3.10$&$0.94$&$0.94$&$1.12$&$1.32$&	$1.19$\\
		\hline
		\hline
	$m_{\Sigma^+}$&$\tau_{B^-}$&$\tau_{B^0}$&$f_{D}$&$f_{D^*}$&$f_{J/\Psi}$&$c_{1}^{D^{(*)}}$&$c_{2}^{D^{(*)}}$&$c_1^{J/\Psi}$&$c_2^{J/\Psi}$\\
		\hline
	$1.19$&$164$&$152$&$0.22$&$0.23$&$0.41$&$-0.367$&$1.169$&$-0.185$&$1.082$\\
		\hline
	\end{tabular}
		\label{input}
\end{table}
By performing the minimum $\chi^2$ method with six data points, the free parameters 
of $C_{RR,LL,LR}$ in Eq.~(\ref{relation}) and the effective color number of $N_c^{eff}$
are fitted to be 
\begin{eqnarray}
	(C_{RR},C_{LL},C_{LR})&=&(-11.67\pm1.97,17.78\pm0.83,6.41\pm1.62)\text{ GeV}^4\,,
	\nonumber\\
	N_c^{eff}&=&0.51\pm0.03\,,
\end{eqnarray}
respectively, with $\chi^2/d.o.f=0.28$. Our fitting results along with the input data for the transition-type  three-body decays of $\bar{B}\to {\bf B}\bar{\bf B}' M$ and $B^-\to  p\bar{ p}\mu^- \bar{\nu}_{\mu}$
are presented in Table.~\ref{fitting}. 
\begin{table}[h]
	\caption{Results for the transition-type decays of $\bar{B}\to {\bf B}\bar{\bf B}' M$ and $B^-\to  p\bar{ p}\mu^- \bar{\nu}_{e}$ .}
	\begin{tabular}{c|c|c}
		\hline
		\hline
		Channel& Data & Our Results\\
		\hline
		$10^5{\cal B}(B^-\to \Lambda^0 \bar{p} J/\Psi)$&$1.46\pm0.12$&$1.47\pm0.12$ \\
		$10^5{\cal B}(B^0\to \Lambda^0 \bar{\Lambda}^0 D)$&$1.00\pm0.30$&$1.23\pm0.10$ \\
		$10^5{\cal B}(B^0\to p \bar{p} D)$&$10.4\pm0.70$&$10.42\pm0.28$ \\
		$10^5{\cal B}(B^0\to p \bar{p} D^*)$&$9.9\pm1.1$&$9.04\pm0.49$ \\
		$10^7{\cal B}(B^0\to p \bar{p} J/\Psi)$&$4.50\pm0.60$&$4.83\pm0.34$ \\
		$10^6{\cal B}(B^-\to  p\bar{ p}\mu^- \bar{\nu}_{\mu})$&$5.27\pm0.35$&$5.21\pm0.34$ \\
		\hline
	\end{tabular}
\label{fitting}
\end{table}
 In Table~\ref{prediction}, we show
 our predictions of other four-body $B^-\to {\bf B}\bar{\bf B}' \ell^- \bar{\nu}_{\ell}$ decays. 
 In Tables~\ref{fitting} and \ref{prediction}, we only consider the errors caused by the data inputs and $\chi^2$ fitting, the other uncertainties
are not listed in our results due to the lack of  a comprehensive model to describe $\bar{B} \to {\bf B}{\bf B}'$. 
The most important source of the theoretical errors is the assumption of the heavy quark limit as shown around Eq.~(\ref{HL}) in Appendix. 
Without this assumption, we need to fit totally 10 parameters, which much exceed the number of the current data points. 
However, from the kinematical point of view, we expect that the error from  the heavy quark approximation should be the same order as that  in $\Lambda_c \to \Lambda$  because of the similar mass ratio of
$2m_{\bf B}/m_{\bar B}\simeq m_\Lambda/m_{\Lambda_c}$ between the two types of the channels.  On the other hand, we are confident with our momentum behaviors of the form factors given by the QCD counting rules, which have been used to explain the threshold effects in three-body $\bar{B}\to {\bf B}{\bf B}'M$ decays. Moreover, these momentum behaviors can match the newest $B^-\to p\bar{ p}\mu^+\nu_{\mu}$ differential decay width measured by the LHCb.
 It is interesting to see that the $SU(3)_f$ flavor symmetry guarantees that all observables in $B^- \to \Lambda^0 \bar{\Sigma}^0 \ell^- \bar{\nu}_{\ell}$  
  are the same as those in $B^- \to \Sigma^0 \bar{\Lambda}^0 \ell^- \bar{\nu}_{\ell}$.
We note that the angular distribution asymmetries in $B^- \to {\bf B}\bar{\bf B}' \ell^- \bar{\nu}_{\ell}$ mainly depend on the electroweak coefficients,
 which are associated with the spin-flavor structures of the ${\bf B}\bar{\bf B}'$ pairs. 
 Interestingly, the angular asymmetries of $B^-\to \Lambda \bar{\Lambda} \ell^- \bar{\nu}_{\ell}$   vanish because only the chiral-conserving interaction
 participates in  $B^-\to \Lambda\bar{\Lambda}$.
As a result, the physical observables in $B^-\to \Lambda \bar{\Lambda} \ell^- \bar{\nu}_{\ell}$ are sensitive to test the availability of pQCD counting rules as well as the asymptotic relations in the limit of $t\to \infty$.
\begin{table}[t]
	\caption{Our numerical results of four-body $B^- \to {\bf B}\bar{\bf B}' \ell^- \bar{\nu}_{\ell}$  decays,
	 where the errors come from the $\chi^2$.}
	\begin{tabular}{c|c|c|c}
		\hline
		\hline
		Channel&$10^6{\cal B}$&$10^2\langle \alpha_{\theta_{\bf B}}\rangle$&$10^2\langle \alpha_{\theta_{\ell}}\rangle$\\
		\hline
		$B^- \to p \bar{p} \ell^- \bar{\nu}_{\ell}$&$5.21\pm0.34$&$-6.51\pm1.51$&$-2.74\pm0.40$\\
		$B^- \to n \bar{n} \ell^- \bar{\nu}_{\ell}$&$0.68\pm0.10$&$4.42\pm1.66$&$0.41\pm0.95$\\
		$B^- \to \Lambda^0 \bar{\Lambda}^0 \ell^- \bar{\nu}_{\ell}$&$0.08\pm0.01$&$0.00$&$0.00$\\
		$B^- \to \Sigma^+ \bar{\Sigma}^ +\ell^- \bar{\nu}_{\ell}$&$0.24\pm0.02$&$-6.91\pm1.62$&$-2.83\pm0.50$\\
		$B^- \to \Sigma^0 \bar{\Sigma}^0 \ell^- \bar{\nu}_{\ell}$&$0.06\pm0.01$&$-6.91\pm1.62$&$-2.83\pm0.49$\\
		$B^- \to \Xi^0 \bar{\Xi}^0 \ell^- \bar{\nu}_{\ell}$&$0.008\pm0.001$&$4.82\pm1.85$&$0.28\pm0.81$\\
		$B^- \to \Lambda^0 \bar{\Sigma}^0 \ell^- \bar{\nu}_{\ell}$&$0.014\pm0.004$&$-5.65\pm2.05$&$-7.88\pm0.64$\\
		$B^- \to \Sigma^0 \bar{\Lambda}^0 \ell^- \bar{\nu}_{\ell}$&$0.014\pm0.004$&$-5.65\pm2.05$&$-7.88\pm0.64$\\
		\hline
	\end{tabular}
	\label{prediction}
\end{table}

\begin{figure}[h]
	\begin{minipage}[h]{0.45\textwidth}
		\centering
		\includegraphics[width=3in]{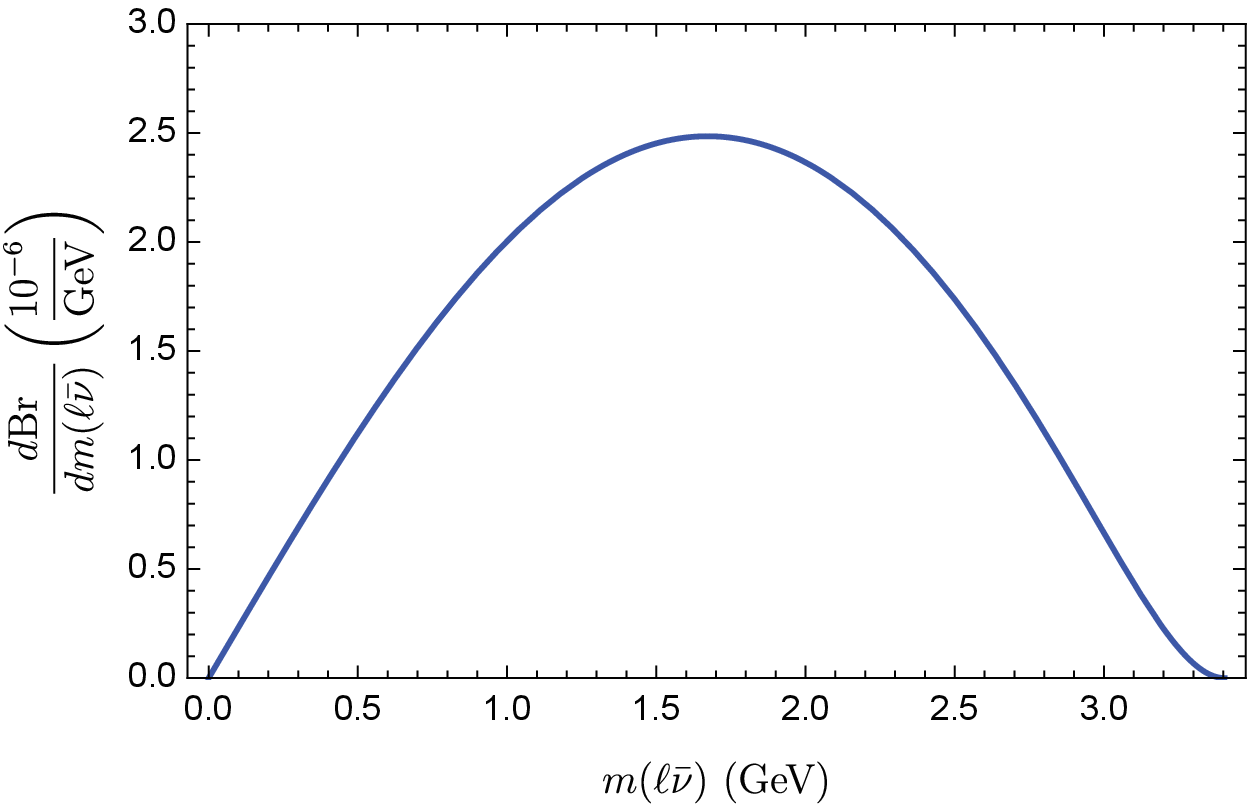}.
	\end{minipage}
\hfill
	\begin{minipage}[h]{0.45\textwidth}
		\centering
		\includegraphics[width=3in]{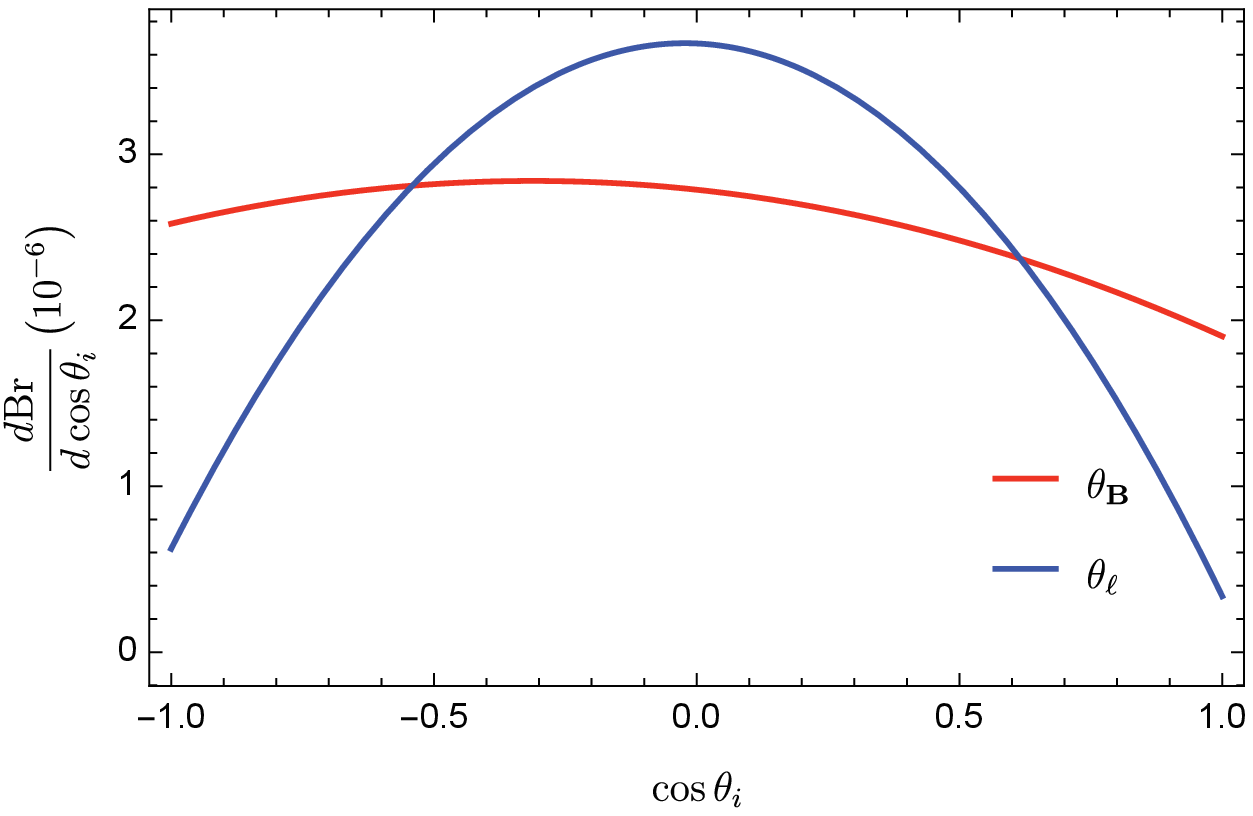}
	\end{minipage}
\caption{Differential branching fractions of $B^- \to p\bar{p} \ell^- \bar{\nu}_{\ell}$
as  functions of  $\ell \bar{\nu}$ invariant mass, $\cos \theta_{\bf B}$ and $\cos \theta_{\ell}$, respectively.}
\label{Bpp}
\end{figure}

\begin{table}[h]
	\caption{Our results  of $B^- \to p \bar{p}\ell^- \bar{\nu}_{\ell}$ along with  those in Ref.~\cite{Geng:2011tr} and the data.} 
	\begin{tabular}{c|c|c|c}
		\hline
		\hline
		&$10^6{\cal B}$&$10^2\langle \alpha_{\theta_{\bf B}} \rangle$&$10^2\langle \alpha_{\theta_{\ell}} \rangle$	\\
		\hline
		Our results&$5.21\pm0.34$&$-6.51\pm1.51$&$-2.74\pm0.40$ \\
		Ref.~\cite{Geng:2011tr}&$104\pm29$&$6\pm2$&$59\pm2$\\
		LHCb ($\ell=\mu$)~\cite{Aaij:2019bdu}
		&$5.27^{+0.23}_{-0.24}\pm0.21\pm0.15$&-&-\\
		Belle ($\ell=e$)~\cite{Tien:2013nga}
		&$8.2^{+3.7}_{-3.2}\pm0.6$&-&-\\
		Belle ($\ell=\mu$)~\cite{Tien:2013nga}
		&$3.1^{+3.1}_{-2.4}\pm0.7$&-&-\\
		Belle (Combined)~\cite{Tien:2013nga} &$5.8^{+2.4}_{-2.1}\pm0.9$&-&-\\
		\hline
	\end{tabular}
\label{Bpplv}
\end{table}
\begin{figure}
	\includegraphics[width=4in]{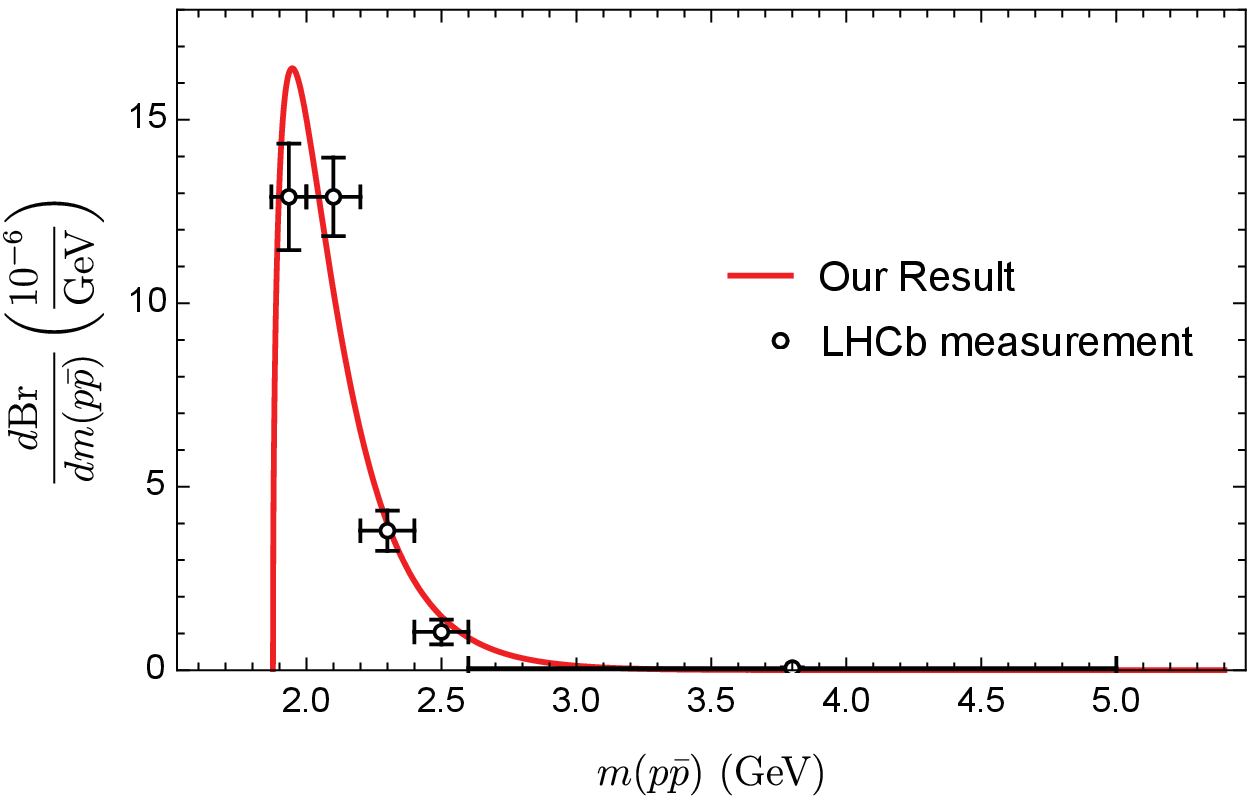}
\caption{Differential branching fraction of $B^-\to p\bar{p}\ell^-\bar{\nu}_{\ell}$ as a function of the $p\bar{p}$ invariant mass ($m(p\bar{p}$),
where the red-solid line is our results, and the hollow dots are the data from the LHCb measurements~\cite{Aaij:2019bdu}.}
\label{com}
\end{figure}
 In Table~\ref{Bpplv}, we summarize our results  of $B^- \to p \bar{p}\ell^- \bar{\nu}_{\ell}$ 
 along with the previous theoretical ones~\cite{Geng:2011tr} as well as
 the experimental data~\cite{Aaij:2019bdu,Tien:2013nga}. We note that the theoretical calculations are insensitive to the lepton mass for the $\ell=e$ and
 $\mu$ channels.
 As seen from Table~\ref{Bpplv}, our new result of  ${\cal B}(B^- \to p \bar{p}\ell^- \bar{\nu}_{\ell})=(5.21\pm0.34)\times 10^{-6}$ 
 is about one order of magnitude lower than the previous theoretical prediction of $(10.4\pm2.9)\times 10^{-5}$ in Ref.~\cite{Geng:2011tr}, but the effective color number in $\bar{B} \to {\bf B}{\bf B}' M$ channel have a magnificent change.
 Our fitting result is consistent  with the Belle measurements of  ${\cal B}(B^- \to p \bar{p}e^- \bar{\nu}_e)=(8.2^{+3.8}_{-3.3})\times 10^{-6}$~\cite{Tien:2013nga}  and 
   ${\cal B}(B^- \to p \bar{p}\mu^- \bar{\nu}_{\mu})=(3.1^{+3.2}_{-2.5})\times 10^{-6}$~\cite{Tien:2013nga}
   and agrees well with the  combined one of $(5.8^{+2.6}_{-2.3})\times 10^{-6}$ by Belle~\cite{Tien:2013nga}
   as well as the recent  $\mu$-channel data of  $(5.3\pm0.4)\times 10^{-6}$ by LHCb~\cite{Aaij:2019bdu} which is one of our input channels. 
   Clearly, more precise modeling and explanation are needed to find the QCD origin of the effective color number being $N_c^{eff}=0.51 \pm 0.03$,  indicating that the non-peturbative effects in three-body $\bar{B}\to {\bf B}{\bf B}' M$ channels is much stronger than those in two-body $\bar{B} \to M_1 M_2$ 
   ones.
  In Fig.~\ref{com}, we plot the $p \bar{p}$ invariant mass spectrum in  $B^-\to p\bar{p}\ell^-\bar{\nu}_{\ell}$.
 By comparing our results  with the LHCb measurement~\cite{Aaij:2019bdu},
   we find that our spectrum is consistent with the observed one. 
We further show the differential branching fractions of  $B^-\to p\bar{p}\ell^-\bar{\nu}_{\ell}$
as  functions of $\sqrt{s}\equiv m_{\ell \bar{\nu}},~\cos_{\theta_{\bf B}}$ and $\cos_{\theta_{\ell}}$ in Fig.~\ref{Bpp}, respectively,
 which can provide us not only the information of the leptonic sector but also the the spin-flavor relations in the $t\to \infty$ asymptotic limit.
 These differential branching fractions could be tested by the ongoing experiments.

\section{Conclusions}
We have systematically revisited the baryonic four-body semileptonic decays of $B^- \to {\bf B} \bar{\bf B}' \ell^- \bar{\nu}_{\ell}$ with $\ell= e,\mu$. 
We have reduced  the  ten  form factors in the hadronic transition of  $B^-\to {\bf B}\bar{\bf B}' $ into three free parameters
in the heavy quark limit and $t\to \infty$.
We have performed the minimum $\chi^2$ method to fit the three parameters and the effective color number of $N_c^{eff}$ with  $\chi^2/d.o.f=0.28$
by using five three-body decays of $\bar{B}\to {\bf B}\bar{\bf B }' M$ along with  the $B^- \to p \bar{ p} \mu^-\bar{\nu}_{\mu}$ measurement.
We have obtained a consistent fitting  result of  ${\cal B}(B^- \to p\bar{p} \ell^- \bar{\nu}_{\ell})=(5.21\pm0.34)\times10^{-6}$ as well as other input channels.
Our $B^- \to p\bar{p} \ell^- \bar{\nu}_{\ell}$ decay branching ratio is about one order of magnitude lower than the previous theoretical 
 prediction  of $(10.4\pm2.9)\times10^{-5}$ in Ref.~\cite{Geng:2011tr}, and
 agrees well with the experimental data of $(5.8^{+2.6}_{-2.3})\times 10^{-6}$ and
$(5.3\pm0.4)\times 10^{-6}$ by Belle~\cite{Tien:2013nga} and LHCb~\cite{Aaij:2019bdu}, respectively.
 In addition, our evaluation of the $m_{p\bar{p}}$ invariant mass spectrum is also consistent with 
  that by the LHCb measurement~\cite{Aaij:2019bdu},  
  demonstrating  that the threshold effect and the $t^{-3}$ dependence of the form factors from  the QCD counting rules is still dominant in the
  baryonic four-body semi-leptonic decays, while the other Lorentz invariant variables, such as $(p_{\bar B}+p_{{\bf B}})^2$, as well as the resonant states are highly suppressed. 
    Furthermore, we have  plotted the differential branching fractions with respect to the kinematic variables
  of $m_{\ell\bar{\nu}}$ and $\cos\theta_{{\bf B},\ell}$ in $B^- \to p\bar{p}\ell^- \bar{\nu}_{\ell}$
  to  provide the information in the lepton sector and angular distributions, respectively. 
  We have also used the flavor symmetry to explore the physical observables in other $B^- \to {\bf B} \bar{\bf B}' \ell^- \bar{\nu}_{\ell}$ decays.
   In particular, 
   we have found that  that the angular asymmetries of $B^- \to \Lambda \bar{\Lambda} \ell^- \bar{\nu}_{\ell}$  vanish 
   due to the absence of the chiral-flipping interaction ($e_{LR}=0$).  On the theoretical side, the non observed transition modes of the
   three-body ${\bar B}\to {\bf B}{\bf B}' M$ and four-body $\bar{B} \to {\bf B}{\bf B}' \ell \bar{\nu}$ decays can help us to relax 
   the assumption of the heavy quark limit once they are measured. Otherwise, the lattice simulation would be currently the most trustworthy theoretical method to reliably extract the hadronic form factors.  On the experimental side, some of our results in $B^- \to {\bf B} \bar{\bf B}' \ell^- \bar{\nu}_{\ell}$
  can be tested by the ongoing experiments at Belle-II and LHCb. 
  Finally, we remark that the theoretical determination of the four-body decays of ${\bar B} \to {\bf B}{\bf B}' \ell \bar{\nu}$ would
  provide a valuable opportunity to  search for  T violating effects from the  triple momentum correlations and improve the measurement of $|V_{ub}|$ as the works in exclusive $B$ and $\Lambda_b$ decays.

\section*{Appendix}
Starting with transition matrix elements of $\langle {\bf B}\bar{\bf B}'|J^{\mu}_{V}-J^\mu_{A}|\bar{B}\rangle$ with $J_{V(A)}^{\mu}=\bar{q}\gamma^{\mu}(\gamma^5)b$, we assume that the $\bar{B}$ meson state can be approximately expressed by the field operator of free quarks $|\bar{B}\rangle\sim \bar{b}\gamma^5q'|0\rangle$. Therefore, the matrix elements become
\begin{eqnarray}
	\langle {\bf B}\bar{\bf B}'|J^{\mu}_{V}-J^\mu_{A}|\bar{B}(\bar{q}'b)\rangle&\simeq&\langle {\bf B}\bar{\bf B}'|\bar{q}\gamma^\mu(1-\gamma^5)b\bar{b}\gamma^5q'|0 \rangle\nonumber\\
	&=&\langle {\bf B}\bar{\bf B}'|\bar{q}\gamma^\mu(1-\gamma^5)(\slashed{p}_{b}+m_b)\gamma^5q'|0 \rangle\nonumber\\
	&=&\langle {\bf B}\bar{\bf B}'|\bar{q}\gamma^\mu(1-\gamma^5)(\slashed{p}_{b}-m_b)q'|0 \rangle\nonumber\\
	 &=&2\langle {\bf B}\bar{\bf B}'|\bar{q}_L\gamma^\mu\slashed{p}_{b}q_R'|0 \rangle-2m_b\langle {\bf B}\bar{\bf B}'|\bar{q}_L\gamma^\mu q_L'|0 \rangle \nonumber \\
	&=&\langle {\bf B}\bar{\bf B}'|J'^\mu|0\rangle-\langle {\bf B}\bar{\bf B}'|\tilde{J}^\mu|0\rangle\,,
\end{eqnarray}
where $J'^{\mu}=2\bar{q}_L\gamma^\mu \slashed{p}_b q_R '$ and  $\tilde{J}^\mu=2m_b\bar{q}_L\gamma^\mu q_L '$ 
with $q_{L(R)}=(1\mp\gamma^5)/2q$ and $\bar{q}_{L(R)}=\bar{q}(1\pm \gamma^5)/2$. 
Note that, by inserting QCD (gluon-quark-antiquark) vertices in the corresponding diagrams, the Dirac structure in Eq.~(11) could be altered. However, because of  the asymptotic freedom in  QCD, we can treat these alterations from QCD vertices as small perturbations, which are negligible in 
the limit of $(p_{\bf B}+p_{{\bf B}'})^2 \to \infty$.
In terms of the crossing-symmetry (c.s.), the final state anti-baryon ($\bar{\bf B}'$) is transformed as the initial baryon (${\bf B}'$) in the initial state with opposite four-momentum $\tilde{p}_{{\bf B}'}=-p_{\bar{\bf B}'}$, resulting in 
\begin{eqnarray}
\langle {\bf B}(p_{\bf B})\bar{\bf B}'(p_{\bar{\bf B}'})|J'^\mu(\tilde{J}^{\mu})|0\rangle \xrightarrow{c.s.}\langle {\bf B}(p_{\bf B})|J'^\mu(\tilde{J}^{\mu})|{\bf B}'(\tilde{p}_{{\bf B}'})\rangle.
\end{eqnarray}
According to Refs.~\cite{Chua:2002yd,Brodsky:1980sx}, 
the amplitude can be parameterized as
\begin{eqnarray}
\langle {\bf B}|J'^\mu|{\bf B}'\rangle&=&2i\bar{u}_{\bf B}\gamma^\mu \slashed{p}_{b}\left( \frac{1+\gamma^5}{2}F'^{+}+\frac{1-\gamma^5}{2}F'^{-}\right)u_{{\bf B}'}\nonumber\\
&=&2i\bar{u}_{\bf B}^L\gamma^\mu \slashed{p}_{b}u_{{\bf B}'}^RF'^{+}+2i\bar{u}_{\bf B}^R\gamma^\mu \slashed{p}_{b}u_{{\bf B}'}^LF'^{-}\nonumber\\
\langle {\bf B}|\tilde{J}^{\mu}|{\bf B}'\rangle&=&2m_bi\bar{u}_{\bf B}\gamma^\mu \left( \frac{1+\gamma^5}{2}\tilde{F}^{+}+\frac{1-\gamma^5}{2}\tilde{F}^{-}\right)u_{{\bf B}'}\,,\nonumber\\
&=&2m_bi\bar{u}_{\bf B}^R\gamma^\mu u_{{\bf B}'}^R \tilde{F}^{+}+2m_bi\bar{u}_{\bf B}^L\gamma^\mu  u_{{\bf B}'}^L\tilde{F}^{-}\,.
\label{csform}
\end{eqnarray}
In  the asymptotic limit of $(p_{\bf B}-\tilde{p}_{{\bf B}'})^2 \to \infty$, the helicity of a particle can be approximately  treated as  its chirality, so that 
 the amplitudes with a specific chirality can be written as
\begin{eqnarray}
 	\langle {\bf B},L|J'^\mu|{\bf B}',R\rangle&=&2i\bar{u}_{\bf B}^{L}\gamma^\mu \slashed{p}_{b}\left( e_{LR}F_{LR}\right)u_{{\bf B}'}^{R}\nonumber\\
 	\langle {\bf B},R(L)|\tilde{J}^{\mu}|{\bf B}',R(L)\rangle&=&2m_bi\bar{u}_{\bf B}^{R(L)}\gamma^\mu \left( e_{RR(LL)}F_{RR(LL)}\right)u_{{\bf B}'}^{R(L)}\,
 	\label{chiralform}
\end{eqnarray}
where
\begin{eqnarray}
	e_{LR}&=&\langle {\bf B},L|(a^{ L}_{q})^{\dagger} a^{R}_{q'}|{\bf B}',R \rangle,\nonumber\\
	e_{RR}&=&\langle {\bf B},R|(a^{ L}_{q})^{\dagger} a^{L}_{q'}|{\bf B}',R \rangle,\nonumber\\ 
	e_{LL}&=&\langle {\bf B},L|(a^{ L}_{q})^{\dagger} a^{L}_{q'}|{\bf B}',L\rangle \,,
	\label{electroweak}
\end{eqnarray}
with the corresponding particle creation and annihilation operators  $(a_{q}^{s})^{\dagger}$  and $a_{q}^{s}$,
and other combinations are zero due to the angular-momentum conservation. 
From Eqs.~(\ref{csform}), (\ref{chiralform}) and (\ref{electroweak}), we find 
that
\begin{eqnarray}
	F'^+=e_{LR}F_{LR},\quad F'^-=0,\quad\tilde{F}^+=e_{RR}F_{RR},\quad \tilde{F}^-=e_{LL}F_{LL}\,.
\end{eqnarray}
Consequently, the transition amplitude in $(p_{\bf B}-\tilde{p}_{{\bf B}'})^2 \to \infty$  is given by
\begin{eqnarray}
\langle {\bf B}|J'^\mu-\tilde{J}^\mu|{\bf B}'\rangle=2i\bar{u}_{\bf B}\gamma^\mu \left(\slashed{p}_{b} \frac{1+\gamma^5}{2}e_{LR}F_{LR}-m_b\frac{1+\gamma^5}{2}e_{RR}F_{RR}-m_b\frac{1-\gamma^5}{2}e_{LL}F_{LL}\right)u_{{\bf B}'}\,.\nonumber \\
\end{eqnarray}
After applying the crossing-symmetry again, we get that
\begin{eqnarray}
\langle {\bf B}\bar{\bf B}'|J^{\mu}_{V}|\bar{B}\rangle&=&i\bar{u}_{\bf B}\gamma^\mu\left(\slashed{p}_b e_{LR}F_{LR}+m_b(e_{LL}F_{LL}-e_{RR}F_{RR})\right)\gamma^5 v_{\bar{\bf B}'}\nonumber\\
&=&i\bar{u}_{\bf B}\gamma^\mu\left(\slashed{p}_{\bar{B}} e_{LR}F_{LR}+m_{\bar{B}}(e_{LL}F_{LL}-e_{RR}F_{RR})\right)\gamma^5 v_{\bar{\bf B}'}\nonumber\\
\langle {\bf B}\bar{\bf B}'|J^{\mu}_{A}|\bar{B}\rangle&=&i\bar{u}_{\bf B}\gamma^\mu\left(-\slashed{p}_b e_{LR}F_{LR}+m_b(e_{LL}F_{LL}+e_{RR}F_{RR})\right) v_{\bar{\bf B}'}\nonumber \\
&=&i\bar{u}_{\bf B}\gamma^\mu\left(-\slashed{p}_{\bar{B}} e_{LR}F_{LR}+m_{\bar{B}}(e_{LL}F_{LL}+e_{RR}F_{RR})\right) v_{\bar{\bf B}'}\,.
\end{eqnarray}
where  we have used the approximations of $p_b\simeq p_{\bar{B}}$ and $m_b\simeq m_{\bar{B}}$ in the heavy quark limit.
With the help of equation of motions
\begin{eqnarray}
	\label{HL}
	\bar{u}_{\bf B}\slashed{p}_{\bf B}=\bar{u}_{\bf B}m_{\bf B},\quad \slashed{p}_{\bar{\bf B}'}v_{\bar{\bf B}'}=-m_{\bar{\bf B}'}v_{\bar{\bf B}'}\,,
	\end{eqnarray}
	 and the Dirac algebra,
	\begin{eqnarray}
	\gamma^{\mu}\slashed{p}_{\bar{B}}&=&\gamma^{\mu}\slashed{p}+\gamma^{\mu}(\slashed{p}_{\bf B}+\slashed{p}_{\bar{\bf B}'})=p^{\mu}-i\sigma^{\mu \nu}p_{\nu}+2p_{\bf B}^\mu-\slashed{p}_{\bf B}\gamma^\mu+\gamma^\mu \slashed{p}_{\bar{\bf B}'}\nonumber \\
	&=&-\slashed{p}_{\bf B}\gamma^\mu+\gamma^\mu \slashed{p}_{\bar{\bf B}'}-i\sigma^{\mu \nu}p_{\nu}+p^{\mu}+(p_{\bf B}-p_{{\bf B}'})^\mu+(p_{\bf B}+p_{\bar{\bf B}'})^{\mu}\,,
\end{eqnarray}
we finally obtain
\begin{eqnarray}
\langle {\bf B}\bar{\bf B}'|J^{\mu}_{V}|\bar{B}\rangle
&=&i\bar{u}_{\bf B}\gamma^\mu\left[-(m_{\bf B}-m_{\bar{\bf B}'}) e_{LR}F_{LR}+m_{\bar{B}}(e_{LL}F_{LL}-e_{RR}F_{RR})\right.\nonumber\\
&+& \left. e_{LR}F_{LR}(-i\sigma^{\mu \nu}p_{\nu}+p^{\mu}+(p_{\bf B}-p_{{\bf B}'})^\mu+(p_{\bf B}+p_{\bar{\bf B}'})^{\mu})\right ]\gamma^5 v_{\bar{\bf B}'}\,,\nonumber\\
\langle {\bf B}\bar{\bf B}'|J^{\mu}_{A}|\bar{B}\rangle
&=&i\bar{u}_{\bf B}\gamma^\mu\left[(m_{\bf B}+m_{\bar{\bf B}'}) e_{LR}F_{LR}+m_{\bar{B}}(e_{LL}F_{LL}-e_{RR}F_{RR})\right.\nonumber\\
&-& \left. e_{LR}F_{LR}(-i\sigma^{\mu \nu}p_{\nu}+p^{\mu}+(p_{\bf B}-p_{{\bf B}'})^\mu+(p_{\bf B}+p_{\bar{\bf B}'})^{\mu})\right ] v_{\bar{\bf B}'}\,,
\end{eqnarray}
which clearly lead to
\begin{eqnarray}
	f_1&=&m_{\bar{B}}\left(e_{LL}F_{LL}+e_{RR}F_{RR}\right)+\left(m_{\bf B}+m_{\bar{\bf B}'}\right)e_{LR}F_{LR}\,,\nonumber\\
	g_1&=&m_{\bar{B}}\left(e_{LL}F_{LL}-e_{RR}F_{RR}\right)-\left(m_{\bf B}-m_{\bar{\bf B}'}\right)e_{LR}F_{LR}\,,\nonumber\\
	f_2&=&-g_2=e_{LR}F_{LR},\quad f_i=-g_i=-e_{LR}F_{LR}\,,
	\quad (i=3,4,5)\,.
	\label{frelation}
\end{eqnarray}
As a result, the constant parts of form factors, $C_{f_i(g_i)}$, $C_{RR(LL)}$ and $C_{LR}$, in Eq.~(\ref{frelation}) 
directly imply the relations in Eq.~(\ref{relation}).

\section*{ACKNOWLEDGMENTS}
This work was supported in part by MoST (MoST-107-2119-M-007-013-MY3).

\end{document}